\documentstyle[12pt]{article}
\input epsf
\begin{document}{\large \bf  ANYONIC BEHAVIOR OF QUANTUM GROUP \hspace*{.15in} FERMIONIC 
AND BOSONIC SYSTEMS}\footnote{Talk given at International Conference on Orbis Scientae 1997, January 23-26
 Miami, Florida}
 \\ \\ \\\hspace*{1in} Marcelo R. Ubriaco\footnote{ubriaco@ltp.upr.clu.edu}\\
\\
\hspace*{1in} Laboratory of Theoretical Physics\\
\hspace*{1.05in}Department of Physics\\
\hspace*{1.05in}University of Puerto Rico\\
\hspace*{1in} R\'{\i}o Piedras Campus, P. O. Box 23343\\
\hspace*{1in} San Juan
PR 00931-3343, USA
\vspace*{.25in}
\baselineskip18pt
\section*{Introduction}

The role of  quantum groups and quantum Lie algebras \cite{Jimbo} in physics
 has its origin in the theory of vertex models \cite{SZ} and the quantum inverse scattering method
\cite{F}.  From the mathematical point of view, two of
the most important developments have been their understanding in terms of the theory
of noncommutative  Hopf algebras \cite{D} and their relation to non-commutative geometry
 \cite{Wo,Manin,WZ}. 

In recent years the study of quantum groups and quantum 
algebras has greatly diversified into several areas of theoretical 
physics. Based on quantum group ideas,
a considerable amount of work was devoted towards a
formulation of the so called $q$-deformed physical systems.
These approaches are attempts to develop more general
formulations of quantum mechanics \cite{U2} and field theory \cite{U3,Su}.
  The main motivation behind this type of projects resides in
searching for new roles that quantum
groups could play in physics other than the theory of
integrable models. A successful and consistent formulation of
a theory involving quantum group symmetries will have the
potential of having new features no present in the standard 
$q\rightarrow 1$ case. Besides, it will provide a more general, or
alternative,  framework to explain physical phenomena.

In this article we show the role that quantum group symmetries,
in particular $SU_q(2)$, play in a thermodynamic
system at high temperatures.  We first
 display the quantum group covariant algebras, which will be used
to build quantum group invariant hamiltonians, and then
we will discuss the behavior of the corresponding 
 quantum group gases at high temperatures,
and show how the parameter $q$ interpolates between a wide range of
attractive and repulsive systems. 

\section*{Quantum Group Covariant Algebras} \label{QGCA}

As it is well known, boson and fermions operators satisfy 
\begin{eqnarray}
\phi_i\phi_j^{\dagger}-\phi_j^{\dagger}\phi_i&=&\delta_{ij}\nonumber\\
\psi_i\psi_j^{\dagger}+\psi_j^{\dagger}\psi_i&=&\delta_{ij}, \label{BF}
\end{eqnarray}
which, for $i,j=1,...N
$, are covariant under $SU(N)$ transformations. 
For the case of unitary quantum group matrices $T$
the coefficients  do not commute but
satisfy for $N=2$ the following algebraic relations
\begin{eqnarray}
T&=&\left(\begin{array}{cc} 
a & b \\ c & d\end{array}\right)\\ 
ab=q^{-1}ba  & , & ac=q^{-1}ca \nonumber \\
bc=cb & , & dc=qcd  \nonumber \\
db=qbd & , &  da-ad=(q-q^{-1})bc  \nonumber \\
& & det_{q}T\equiv ad-q^{-1}bc=1 ,
\end{eqnarray} 
with the unitary condition \cite{VZW} $\overline{a}=d, \overline{b}=q^{-1}c$
and $q\in {\bf R}$. Hereafter, we take $0\leq q<\infty$.

 A natural question
to address is which are the quantum group
analogues of Equation (\ref{BF}), which will tell us for example
how to build quantum group invariant hamiltonians. The operator
algebras covariant under the action of $SU_q(N)$ matrices
were given in \cite{U4}
\begin{equation}
\Omega_j\overline{\Omega}_i=\delta_{ij}\pm q^{\pm1}R_{kijl}
\overline{\Omega}_l\Omega_k \label{c1}
\end{equation}
\begin{equation}
\Omega_l\Omega_k=\pm q^{\mp 1}R_{jikl}\Omega_j\Omega_i,\label{c2}
\end{equation}
where $\Omega=\Phi,\Psi$ and the upper (lower) sign applies to quantum group bosons $\Phi_i$
(quantum group fermions $\Psi_i$) operators.  The $N^2\times N^2$ matrix 
 $R_{jikl}$ is explicitly written as \cite{WZ}
\begin{equation}
R_{jikl}=\delta_{jk}\delta_{il}(1+(q-1)\delta_{ij})
+(q-q^{-1})\delta_{ik}\delta_{jl}\theta(j-i),
\end{equation}
where $\theta(j-i)=1$ for $j>i$ and zero otherwise. Denoting 
the new fields as $\Omega_i'=\sum_{i=1}
^{N}T_{ij}\Omega_j$, the
$SU_{q}(N)$ transformation matrix $T$ 
and the $R$-matrix  
satisfy the well known algebraic relations \cite{Ta}
\begin{equation}
RT_1T_2=T_2T_1R,\label{T}
\end{equation}
and
\begin{equation}
R_{12}R_{13}R_{23}=R_{23}R_{13}R_{12},
\end{equation}
with the standard embedding $T_1=T\otimes 1$, $T_2=1\otimes T$
$\in V\otimes V$ and $(R_{23})_{ijk,i'j'k'}=
\delta_{ii'} R_{jk,j'k'} \in V\otimes V\otimes V$.

In particular, for $N=2$, Equations (\ref{c1}) and (\ref{c2}) 
are simply written
\begin{description}
\item[a)] $SU_q(2)-fermions$
\begin{eqnarray}
\{\Psi_{2},\overline{\Psi}_{2}\}&=&1\\ \label{f0}
\{\Psi_{1},\overline{\Psi}_{1}\}&=&1 - (1-q^{-2})\overline{\Psi}_{2}\Psi_{2}\label{f1}\\ 
\Psi_{1}\Psi_{2}&=&-q \Psi_{2}\Psi_{1}\\ 
\overline{\Psi}_{1}\Psi_{2}&=&-q \Psi_{2}\overline{\Psi}_{1}\\
\{\Psi_{1},\Psi_{1}\}&=&0=\{\Psi_{2},\Psi_{2}\} \label{0},
\end{eqnarray}
\item[b)] $SU_q(2)-bosons$
\begin{eqnarray}
\Phi_2\overline{\Phi}_2-q^2\overline{\Phi}_2\Phi_2&=&1 \label{b1}\\
\Phi_1\overline{\Phi}_1-q^2\overline{\Phi}_1\Phi_1&=&1+(q^2-1)\overline{\Phi}_2\Phi_2
\\
\Phi_2\Phi_1&=&q\Phi_1\Phi_2\\
\Phi_2\overline{\Phi}_1&=&q\overline{\Phi}_1\Phi_2,\label{b4}
\end{eqnarray}
\end{description}
which for $q=1$ become the fermion and boson algebras respectively.  These operator relations
are very different than those satisfied by the so called $q$-fermions \cite{NG} and
$q$-bosons \cite{Mf,B}, which are written respectively as
\begin{description}
\item[c)] $q$-fermions
\begin{eqnarray}
bb^\dagger+qb^\dagger b&=&q^{N}\label{qf}\\
b^\dagger b&=&[N]\\
bb^\dagger&=&[1-N]\\
b^2=&0&=b^{\dagger 2}, 
\end{eqnarray}
where the bracket $[x]=\frac{q^x-q^{-x}}{q-q^-1}$.
\item[d)] $q$-bosons
\begin{equation}
a_i a_i^{\dagger}-q^{-1}a_i^{\dagger}a_i=q^N ,\;\;\;
[a_i,a_j^{\dagger}]=0=[a_i,a_j],\label{qb}.
\end{equation}
\end{description}
It is simple to check that Equations (\ref{qf})-(\ref{qb})
are not quantum group covariant, and therefore a quantum group
action on the operators $b_i$ and $a_j$ cannot be defined.  Hereafter,
we discuss the thermodynamic properties of the systems described
by the simplest quantum group
invariant hamiltonians.
\newpage
\section
*{Quantum Group Fermion and Boson Models}
{\bf Quantum Group Fermion Gas}\\

From Equation (\ref{0}) we see that for quantum group fermions the occupation
numbers are restricted to $m=0$ or $1$, and therefore $SU_q(N)$-fermions
satisfy the Pauli exclusion principle.  For a given $\kappa$,
a normalized state is simply written as
\begin{equation}
\overline{\Psi}^{n}_{2}\overline{\Psi}^{m}_{1}|0\rangle \;\;\; n,m=0,1,
\end{equation}
and the operator ${\cal M}_i\equiv \overline{\Psi}_i\Psi_i$ satisfy
\begin{equation}
[{\cal M}_2,\Psi_1]=0={\cal M}_1\Psi_2-q^2\Psi_2{\cal M}_1.
\end{equation}
A representation of the $\Psi$ operators in terms of ordinary fermions
$\psi_j$ is simply given by the following relations
\begin{equation}
\Psi_m=\psi_m\prod_{l=m+1}^{N}\left(1+(q^{-1}-1)M_l\right), \label{frep1}
\end{equation}
\begin{equation}
\overline{\Psi}_m=\psi_m^{\dagger}\prod_{l=m+1}^{N}\left(1+(q^{-1}-1)M_l\right), \label{frep2}
\end{equation}
where $M_l=\psi_l^{\dagger}\psi_l$.  

The simplest Hamiltonian one can write in terms of the operators $\Psi_i$ is simply
the one that becomes the free fermion Hamiltonian for $q=1$.  It is given by \cite{U5}
\begin{equation}
{\cal H}_F=\sum_{\kappa}\varepsilon_{\kappa}({\cal M}_{1,\kappa}+{\cal M}_{2,\kappa}),
\end{equation}
where ${\cal M}_{i,\kappa}=\overline{\Psi}_{i,\kappa}\Psi_{i,\kappa}$ and
$\{\overline{\Psi}_{i,\kappa},\Psi_{j,\kappa'}\}=0$ for $\kappa\neq\kappa'$.
With use of the fermion representation in Equations (\ref{frep1}) and (\ref{frep2}), the original 
Hamiltonian becomes the interacting fermion  Hamiltonian
\begin{equation}
{\cal H}_F=\sum_\kappa \varepsilon_\kappa\left(M_{1,\kappa}+M_{2,\kappa}+(q^{-2}-1)
M_{1,\kappa}M_{2,\kappa}\right).\label{HF}
\end{equation}
We see that the parameter $q\neq 1$ mixes the two degrees of freedom in
a nontrivial way through a quartic interaction term.  The grand partition function for
this model is simply written as
\begin{eqnarray}
{\cal Z}_F&=&\prod_\kappa\sum_{n=0}^1\sum_{m=0}^1e^{-\beta\varepsilon_\kappa(n+m
-(1-q^{-2})mn}e^{\beta\mu(n+m)}\\
&=&\prod_\kappa \left(1+2e^{-\beta(\varepsilon_\kappa-\mu)}+e^{-\beta\left
(\varepsilon_\kappa(q^{-2} +1)-2\mu\right)}\right),\label{ZF}
\end{eqnarray}
which for $q=1$ becomes the square of a single-fermion-type grand partition
function.  For a high temperature (or low density) gas, we expand
the grand partition function ${\cal Z}_F$ in terms of the
fugacity $z\ll 1$
\begin{equation}
\ln {\cal Z}_F=4V(2m\pi/h^2\beta)^{3/2}\left[\frac{z}{2}-\alpha(q) \frac{z^{2}}{2}
+\gamma(q)\frac{z^{3}}{3!}+...\right],\label{zf}
\end{equation}
where  the functions $\alpha(q)$ and $\gamma(q)$ are
\begin{eqnarray}
\alpha(q)&=&\frac{1}{2^{3/2}}-\frac{1}{2(q^{-2}+1)^{3/2}}\nonumber\\ 
\gamma(q)&=&\frac{4}{3^{3/2}}-\frac{3}{(q^{-2}+2)^{3/2}}.\nonumber
\end{eqnarray}
Calculating the average number of particles 
$\langle M\rangle=\frac{1}{\beta}\left(\frac{\partial\ln{\cal Z}_F}{\partial\mu}
\right)_{T,V}$ and reverting the equation to write the
fugacity in terms of $\langle M\rangle$ gives for Equation (\ref{zf})
\begin{equation}
\ln{\cal Z}_F=\langle M\rangle\left[1+\frac{\alpha(q)\langle M\rangle}{2V}\lambda_T^3
-\frac{\langle M\rangle^{2}}{16V^{2}}\lambda_T^6\Lambda+...\right],
\end{equation}
where $\Lambda=\frac{8\gamma(q)}{3}+16\alpha^{2}(q)$ and $\lambda_T=\left(h^2\beta/2\pi m\right)^{1/2}$.

From this equation we can obtain the internal energy $U=-\frac{\partial \ln {\cal Z}_F}{\partial\beta}
+\mu\langle M\rangle$, the heat capacity $C_{v}=\left(\frac{\partial U}{\partial T}\right )_V$ and
the entropy $S=\frac{U-\mu\langle M\rangle}{T}+k\ln{\cal Z}_F$ as 
functions of $\langle M\rangle$. The corresponding
equations are
\begin{equation}
U=\frac{3\langle N\rangle}{2\beta}\left[1+\frac{\langle M\rangle}{2V}\lambda_T^3\alpha(q)-\frac{\langle M\rangle^2}
{16V^2}\lambda_T^6\Lambda+...\right]
\end{equation}
\begin{equation}
C_{v}=\frac{3\langle M\rangle k}{2}\left[1-\frac{\langle M\rangle}{4V}
\lambda_T^3\alpha(q)
+\frac{\langle M\rangle^{2}}{8V^2}\lambda_T^6\Lambda+...\right],
\end{equation}
\begin{equation}
S=\langle M\rangle k\left[\frac{5}{2}-\ln\left(\frac{\langle M\rangle}{2V}\lambda_T^3\right)
+\frac{\langle M\rangle}{4V}\lambda_T^3\alpha(q)+...\right].
\end{equation}
The equation of state  is given by the equation
\begin{equation}
pV=kT\langle M\rangle\left[1+\frac{\langle M\rangle}{2V}\lambda_T^3\alpha(q)+...\right].
\end{equation}
Clearly, all these functions become, for $q=1$, the thermodynamic functions for
an ideal fermion gas with two species.  The sign of the second virial coefficient
depends of the value of $q$, implying then that the parameter $q$
interpolates between repulsive and attractive systems.  
Figure
 1 shows a graph of the coefficient $\alpha(q)$. The function $\alpha(q)$ takes
values in the interval $2^{-5/2}\leq\alpha\leq 2^{-3/2}$
for  $0\leq\ q\leq 1$, vanishes at $q=1.96$ and it gets its lowest
 value  $\alpha(q)=-2^{-5/2}(\sqrt{2}-1)$ in the limit $q\rightarrow\infty$.
It is important to remark that  the second virial coefficient for
the ideal boson gas case $B_{bosons}=-2^{-7/2}\beta^{3/2}<B(q\rightarrow\infty,T)=-2^{-5/2}(\sqrt{2}-1)
\beta^{3/2}$, and therefore free bosons are not described in this model.

A natural
question to address is whether a similar interpolation occurs
at $D=2$. Repeating the previous procedure leads to the
equation of state
\begin{equation}
pA=kT\langle M\rangle\left(1+\frac{1}{4(1+q^2)}
\frac{\langle M\rangle}{A}\lambda_T^2+...\right),
\end{equation}
wherein  the second virial coefficient is positive for all values
of $q$, showing  that this model
, at $D=2$, describes only interacting fermionic systems. 

\epsfxsize=475pt \epsfbox{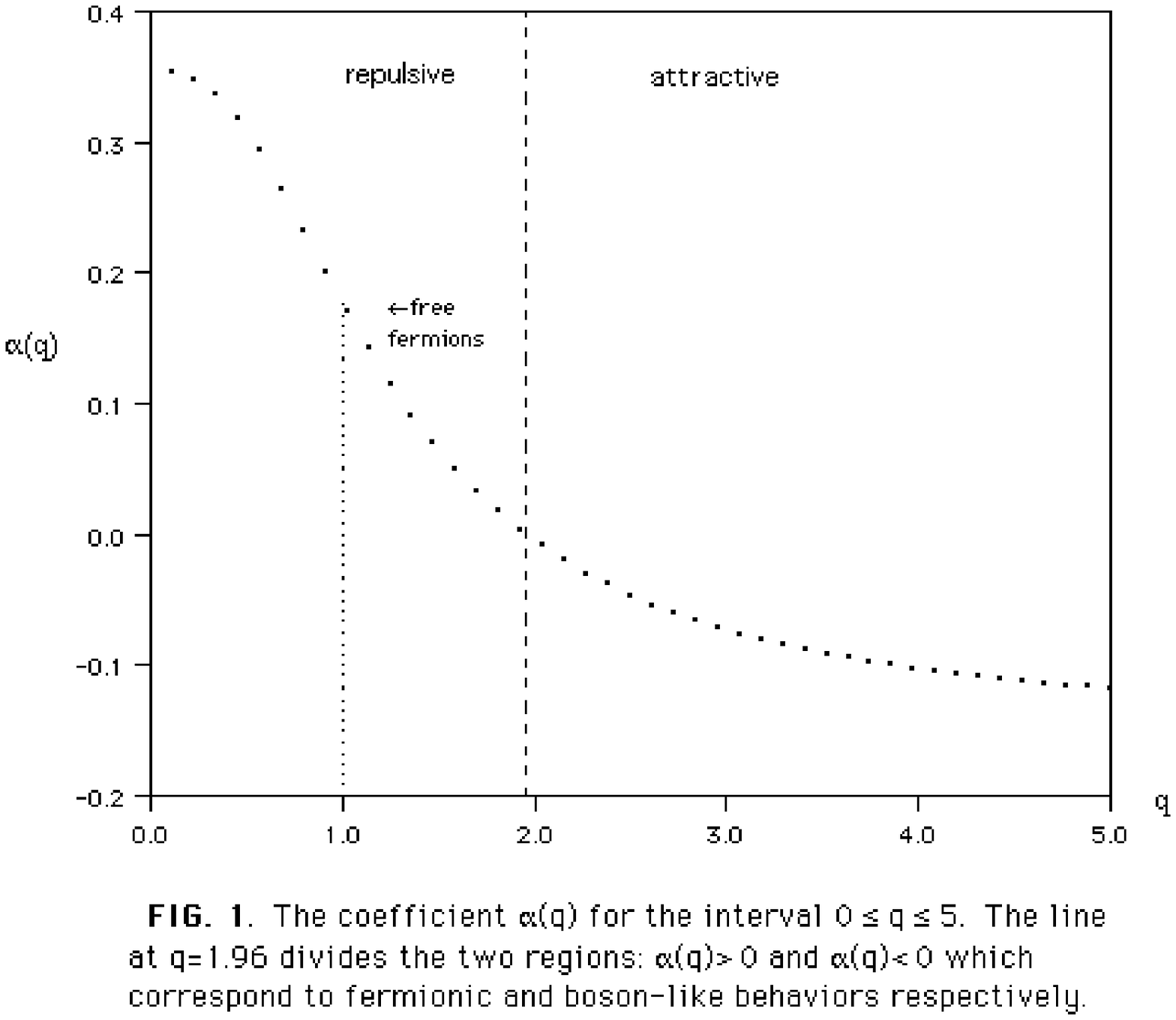}

{\bf Quantum Group Boson Gas}\\

A representation of the quantum group bosons in terms of
 boson operators $\phi_i$ and $\phi^{\dagger}_j$, according
to Equations (\ref{b1})-(\ref{b4}), is simply given by 
\begin{eqnarray}
\Phi_2&=&(\phi_2^\dagger)^{-1} \{N_2\}\label{r1}\\
\overline{\Phi}_2&=&\phi_2^\dagger \\
\Phi_1&=&(\phi_1^\dagger)^{-1} \{N_1\}q^{N_2}\\
\overline{\Phi}_1&=&\phi_1^\dagger q^{N_2}\label{r2},
\end{eqnarray}
where the bracket $\{x\}=\frac{1-q^{2x}}{1-q^2}$ and the boson
number operator $N_i=\phi^{\dagger}_i
\phi_i$.  Therefore,  the simplest quantum group invariant Hamiltonian \cite{U6}
${\cal H}_B$ 
\begin{equation}
{\cal H}_B=\sum_\kappa \varepsilon_\kappa({\cal N}_{1,\kappa}+{\cal N}_{2,\kappa}),
\end{equation}
with $[\overline{\Phi}_{i,\kappa},\Phi_{j,\kappa'}]=0$ for
$\kappa\neq\kappa'$, becomes the interacting bosonic Hamiltonian
\begin{equation}
{\cal H}_B=\sum_{\kappa}\varepsilon_{\kappa}\{\phi_{1,\kappa}^{\dagger}
\phi_{1,\kappa}+\phi_{2,\kappa}^{\dagger}
\phi_{2,\kappa}\},
\end{equation}
with the bracket $\{x\}$ as defined below Equation (\ref{r2}).
Now, it is simple to write the grand partition function ${\cal Z}_B$ for this model.
Introducing the chemical potential $\mu$ in the usual way gives
\begin{equation}
{\cal Z}_B=\prod_\kappa\sum_{n=0}^{\infty}\sum_{m=0}^{\infty}
e^{-\beta\varepsilon_\kappa\{n+m\}}e^{\beta\mu(n+m)},\label{Zb},
\end{equation}
such that after rearrangement of equal power terms it simplifies to
the expression
\begin{equation}
{\cal Z}_B=\prod_\kappa\sum_{m=0}^{\infty} (m+1)e^{-\beta\varepsilon_{\kappa}\{m\}}z^m.
\end{equation}

In $D=3$ the first few terms in powers of $z$ read
\begin{eqnarray}
\ln{\cal Z}_B&=&\frac{4\pi V}{h^3}\int_{0}^{\infty}dp p^2( 2e^{-\beta\varepsilon_{\kappa}}z
+(6 e^{-\beta\varepsilon_{\kappa}\{2\}}-4e^{-\beta\varepsilon_{\kappa}2})\frac{z^2}{2}\nonumber\\
&+&(24 e^{-\beta\varepsilon_{\kappa}\{3\}}-36 e^{-\beta\varepsilon_{\kappa}\{2\}}
e^{-\beta\varepsilon_{\kappa}}
+16 e^{-\beta\varepsilon_{\kappa}3})\frac{z^3}{3!}+...),
\end{eqnarray}
such that performing the elementary integrations gives
\begin{equation}
\ln {\cal Z}_B=\frac{4\pi V}{h^3}\left(\frac{\sqrt{\pi}}{2}(\frac{2m}{\beta})^{3/2} z+
\sqrt{\pi}(\frac{2m}{\beta})^{3/2}\delta(q) z^2+...\right),
\end{equation}
where $\delta(q)=\frac{1}{4}\left(\frac{3}{(1+q^2)^{3/2}}-\frac{1}{\sqrt{2}}\right)$. 

Calculating the
average
 number of particles $\langle N\rangle=\frac{1}{\beta}\left(\frac{\partial\ln{\cal Z}_B}{\partial\mu}\right)
_{T,V}$ and reverting the equation we find for the fugacity 
\begin{equation}
z\approx \frac{1}{2}\left(\frac{h^2}{2m\pi kT}\right)^{3/2}
\frac{\langle N\rangle}{V}-\delta(q) \left(\frac{h^2}{2m\pi kT}\right)^{3}\left(\frac{\langle N
\rangle}{V}\right)^2.
\end{equation}
The internal energy, heat capacity and entropy functions in terms of
the average number of particles $\langle N\rangle$ and $q$ read
\begin{equation}
U=\frac{3\langle N\rangle}{2\beta}\left[1-\frac{\langle N\rangle}{V}\lambda_T^3\delta(q)
+\frac{\langle N\rangle^2}{V^2}\lambda_T^6\left(4\delta^2(q)-\frac{\Gamma(q)}{12}\right)+...\right],
\end{equation}
\begin{equation}
C_v=\frac{3k\langle N\rangle}{2}\left[ 1+\frac{\langle N\rangle}{2V}\lambda_T^3\delta(q)
-2\frac{\langle N\rangle^2}{V^2}\lambda_T^6\left(4\delta^2(q)-\frac{\Gamma(q)}{12}\right)+...\right],
\end{equation}
\begin{equation}
S=k\langle N\rangle\left[\frac{5}{2}-\ln\left(\frac{\langle N\rangle}{2V}\lambda_T^3\right)-
\frac{\langle N\rangle}{2V}\lambda_T^3\delta(q)+...\right],
\end{equation}
where the function $\Gamma(q)=\frac{12}{3^{3/2}}-\frac{18}{(2+q^2)^{3/2}}+\frac{8}{3^{3/2}}$.
The equation of state for this model is more interesting than for the
$SU_q(2)$ fermion gas. For $D=3$, the equation of state is given by
\begin{equation}
pV=kT\langle N\rangle\left(1-\frac{\langle N\rangle}{V}\lambda_T^3\delta(q)
+...\right).
\end{equation}
As expected, at $q=1$ the coefficient $\delta(1)=2^{-7/2}$, which is the numerical
factor in the second virial coefficient for a free boson gas with two species.
The free fermion $\delta(q)=-2^{7/2}$ and ideal gas $\delta(q)=0$
 cases   are reached at  $q\approx 1.78$
and $q\approx 1.27$ respectively. 

A similar calculation for $D=2$ leads to the equation of state
\begin{equation}
pA=kT\langle N\rangle\left(1-\frac{\langle N\rangle}{A}\lambda_T^2\eta(q)+...\right),\label{pA}
\end{equation}
with $\eta(q)=\frac{(2-q^2)}{4(1+q^2)}$. Figure 2 shows a graph of the coefficient $\eta(q)$
 as a function of the parameter
$q$ for $D=2$.  The coefficient $\eta(q)$ in Equation (\ref{pA})
takes values in the interval $[-\frac{1}{4},\frac{1}{2}]$.  At $D=2$ this model
behaves as a fermion gas at $q=\sqrt{5}$.

\epsfxsize=475pt \epsfbox{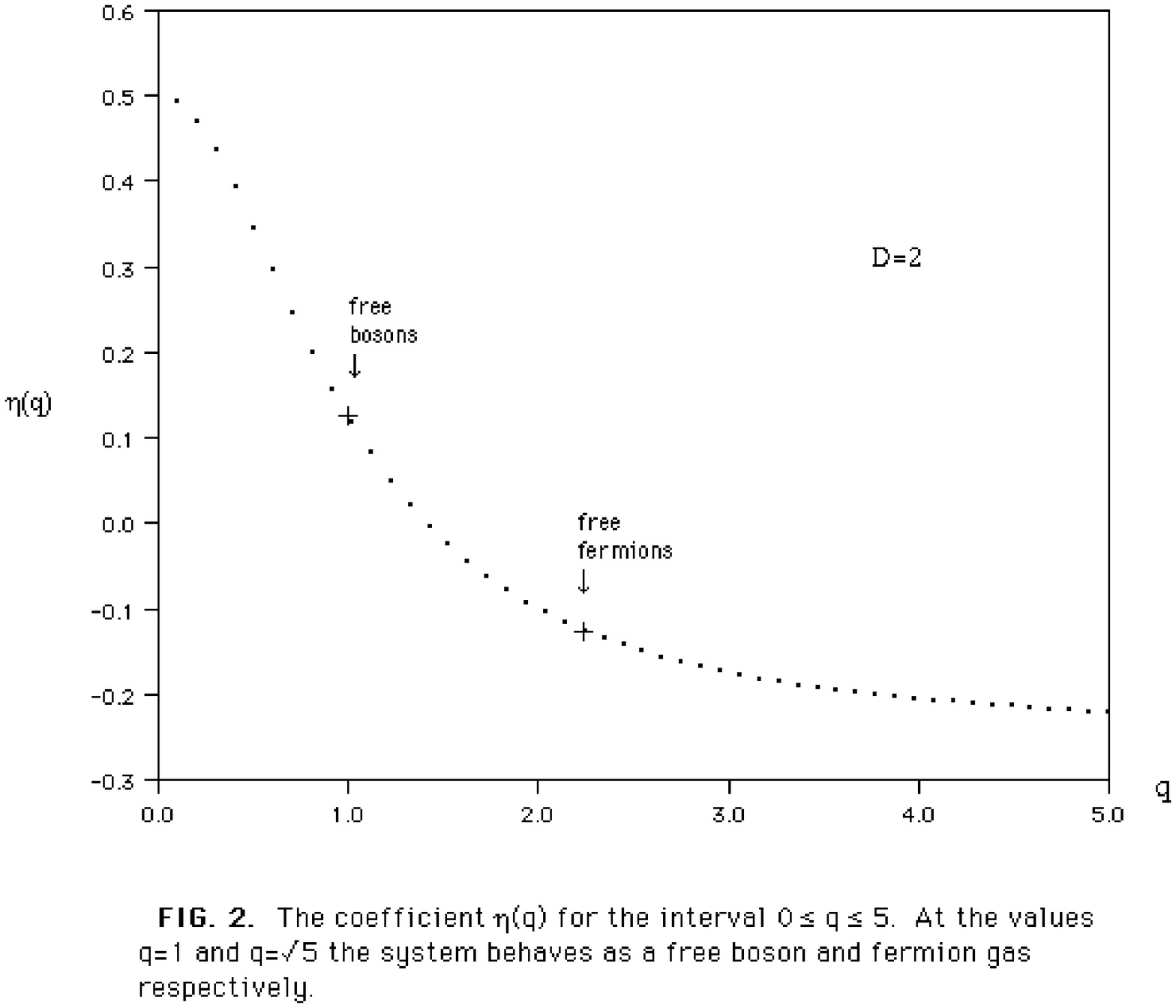}

Since the $SU_q(2)$
boson gas at $D=2$ also interpolates completely between bosons and fermions,
 we can find a relation
between the parameter $q$ and the  statistical parameter $\alpha$ for an
anyon gas \cite{Wi} of two species. This relation is given by
\begin{equation}
\alpha=1-\sqrt{\frac{5-q^2}{2(1+q^2)}},
\end{equation}
where $0\leq\alpha\leq 1$.  The parameter $q$ interpolates within a larger range of
attractive and repulsive systems than the $\alpha$ parameter does.

\section*{Discussion}

In this article we have discussed the high temperature behavior of
quantum group gases.
We considered the two simplest quantum group invariant Hamiltonians
, which are those that become for $q=1$ the free fermion or boson
gases with two species.  A representation of the quantum group fermions in terms
of ordinary fermions  leads 
to a  fermion system with a quartic interaction whose 
coupling constant vanishes as $q\rightarrow 1$.  At high temperatures we 
analyzed
the equation of state at $D=2$ and $D=3$ spatial dimensions.
At $D=2$ the second virial coefficient is always positive for
all values of $q$, therefore in two dimensional space this model describes
only interacting fermion systems. At $D=3$ the sign of the second
virial coefficient depends of the value of $q$, showing then that
the parameter $q$ interpolates between repulsive and attractive
behavior. The ideal gas case corresponds to $q=1.96$ and the system becomes
repulsive for $q<1.96$.  For $q>1.96$ the system becomes attractive, but as
$q\rightarrow\infty$ the free boson limit is not reached, and therefore this model
does not interpolate completely between the free fermion and free boson cases.

For $SU_q(2)$ bosons the results are more interesting. A 
representation of the quantum group boson operators in terms of ordinary bosons leads to a
 hamiltonian in terms of ordinary boson interactions involving powers
of the number operators and $\ln q$. For $D=2$
and $D=3$ the $q$ parameter interpolates completely between the free boson and free 
fermion cases.  For $D=2$, a comparison with the anyon statistical parameter
shows that the parameter $q$ interpolates within a larger range of systems.

Thus, at high temperatures the  interactions that result by imposing
$SU_q(2)$ symmetry in the simplest hamiltonian are such that these models, and
in particular the quantum group boson model, offer
an alternative approach in describing systems obeying fractional statistics
in two and three spatial dimensions.

\end{document}